\documentclass[fleqn,usenatbib]{mnras_tex_edited}

\usepackage{amssymb,amsmath,verbatim,mathtools,needspace,enumitem,etoolbox,graphicx,physics,microtype,natbib,url,hyperref,mathrsfs,bm,ulem,lipsum}
\usepackage{orcidlink}
\usepackage{siunitx}
\newcommand{\ssim}{\mathchar"5218\relax\,}

\renewcommand{\emph}[1]{\textit{#1}}

\newcommand{\approptoinn}[2]{\mathrel{\vcenter{
  \offinterlineskip\halign{\hfil$##$\cr
    #1\propto\cr\noalign{\kern2pt}#1\sim\cr\noalign{\kern-2pt}}}}}

\newcommand{\new}[1]{{#1}}

\def\apj{{Astrophys.~J.}}\def\apjl{{Astrophys.~J.~Lett.}}
\def\apjs{{Astrophys.~J.~Supp.}}
\def\ao{{Appl.~Opt.}}

\def\aap{{Astron.~Astrophys.}}

\def\mnras{{Mon.~Not.~R.~Astron.~Soc.}}

\def\prd{{Phys.~Rev.~D}}

\def\prx{{Phys.~Rev.~X}}
\def\prl{{Phys.~Rev.~Lett.}}
\def\pasa{{Publ. Astron. Soc. Aust.}}

\def\nat{{Nature}}

\def\physrep{{Phys.~Rep.}}

\def\lrr{{Living Rev. Relativ.}}
\def\cqg{{Classical~Quant.~Grav.}}
\def\natastro{{Nat. Astron.}}

\newcommand{\DAMTP}{Department of Applied Mathematics and Theoretical Physics, Cambridge CB3 0WA, United Kingdom}
\newcommand{\SPA}{School of Physics and Astronomy, Monash University, Clayton VIC 3800, Australia}
\newcommand{\OzGravMonash}{OzGrav: The ARC Centre of Excellence for Gravitational Wave Discovery, Clayton VIC 3800, Australia}

\newcommand{\milan}{Dipartimento di Fisica ``G. Occhialini'', Universit\'a degli Studi di Milano-Bicocca, Piazza della Scienza 3, 20126 Milano, Italy}
\newcommand{\infn}{INFN, Sezione di Milano-Bicocca, Piazza della Scienza 3, 20126 Milano, Italy}
\newcommand{\bham}{School of Physics and Astronomy \&	 Institute for Gravitational Wave Astronomy, University of Birmingham,\vspace{-0.05cm}\\$\;$Birmingham, B15 2TT, UK}
\newcommand{\rome}{Dipartimento di Fisica, ``Sapienza'' Universit\`a di Roma \& Sezione INFN Roma1, Piazzale Aldo Moro 5, 00185, Roma, Italy}

\title[Eccentricity or spin precession?]{Eccentricity or spin precession? Distinguishing subdominant effects in gravitational-wave data}

\author[Romero-Shaw, Gerosa, Loutrel]
{Isobel M. Romero-Shaw\thanks{\href{mailto:ir346@cam.ac.uk}{ir346@cam.ac.uk}}$\,$\orcidlink{0000-0002-4181-8090}$^{1,2,3}$,
Davide Gerosa$\,$\orcidlink{0000-0002-0933-3579}$^{4,5,6}$,
Nicholas Loutrel$\,$\orcidlink{0000-0002-1597-3281}$^{7}$
\medskip
\\
$^{1}$\DAMTP\\
$^{2}$\SPA\\
$^{3}$\OzGravMonash\\
$^{4}$\milan\\
$^{5}$\infn\\
$^{6}$\bham\\
$^{7}$\rome
}

\date{}

\pubyear{\the\year{}}

\begin{document}
\label{firstpage}
\pagerange{\pageref{firstpage}--\pageref{lastpage}}
\maketitle

\begin{abstract}
Eccentricity and spin precession are key observables in gravitational-wave astronomy, encoding precious information about the astrophysical formation of compact binaries together with fine details of the relativistic two-body problem. 
However, the two effects can mimic each other in the emitted signals, raising issues around their distinguishability. 
Since inferring the existence of both eccentricity and spin precession simultaneously is---at present---not possible, current state-of-the-art analyses assume that either one of the effects may be present in the data. In such a setup, what are the conditions required for a confident identification of either effect?
We present simulated parameter inference studies in realistic LIGO/Virgo noise, studying events consistent with either spin precessing or eccentric binary black hole coalescences and recovering under the assumption that either of the two effects may be at play.
We quantify how the distinguishability of eccentricity and spin precession increases with the number of visible orbital cycles, confirming that the signal must be sufficiently long for the two effects to be separable. The threshold depends on the injected source, with inclination, eccentricity, and effective spin playing crucial roles.  
In particular, for injections similar to GW190521, we find that it is impossible to confidently distinguish eccentricity from spin precession.
\end{abstract}


\section{Introduction}
\label{sec:intro}

While the masses of merging binary black holes (BBHs) and certain aligned combinations of their spins are now well measured from gravitational waves~\citep[GWs,][]{2019PhRvX...9c1040A,2021arXiv211103606T,2021PhRvX..11b1053A}, subdominant parameters remain relatively elusive. 
The next-in-line targets for GW astronomy are spin precession and orbital eccentricity. 
A confident identification of precessing spins and eccentricity in the events observed by LIGO \citep{2015CQGra..32g4001L} and Virgo \citep{2015CQGra..32b4001A} will not only provide observational constraints on the relativistic dynamics of BBHs, but also constitute a key step in deducing the astrophysical formation mechanisms producing the observed mergers.
While binaries formed in isolation are expected to be observed with negligible orbital eccentricity %
and spins that are closely aligned to the %
orbital angular momentum, those that become bound via dynamical interactions may exhibit a residual detectable eccentricity in band as well as largely misaligned spins (for reviews, see \citealt{2021hgwa.bookE..16M,2022PhR...955....1M}).

Although both eccentricity and spin precession in close-to-merger BBHs are considered signs of dynamical formation, the mechanisms driving the two effects are substantially different. 
In the gravitational two-body problem, eccentricity decays faster than the orbital separation \citep{1964PhRv..136.1224P}. 
A residual eccentricity within the LIGO-Virgo-KAGRA sensitivity band ($\ssim 10$~Hz) therefore indicates that the two black holes (BHs) became bound somewhat recently. 
Predictions for the detectable eccentricity distributions expected from different environments are now available, indicating that large sets of eccentric events could be used to dissect their underlying contribution to the observed merger rate (cf. e.g.~\citealt{2011A&A...527A..70K,2017ApJ...840L..14S,2021ApJ...921L..43Z})

Spin magnitudes %
are largely set by the formation mechanism of each individual BH.
When a BH forms through stellar collapse, its spin crucially depends on the coupling strength between the core and the envelope of the star \citep{2019ApJ...881L...1F,2020A&A...636A.104B}. 
When a BH instead forms as the remnant of a previous BBH merger, conservation of angular momentum through plunge imparts a  dimensionless spin magnitude of $\ssim 0.7$ \citep{2005PhRvL..95l1101P, 2008ApJ...684..822B, 2021NatAs...5..749G}. %
As for the spin directions, these are expected to be randomly distributed for at least some of the dynamically formed systems,  
causing the orbital plane to precess around the total angular momentum of the binary \citep{1994PhRvD..49.6274A}.
Meanwhile, %
binaries formed in isolation share the overall angular momentum of the environment, resulting in BBHs with predominantly aligned spins.  However, some amount of spin precession is also expected for isolated systems because of supernova kicks \citep{2000ApJ...541..319K,2018PhRvD..98h4036G}, potentially polluting the sample of BBHs formed dynamically.

Statistical inference of the source properties of GW events relies on readily available and computationally efficient signal models ---an aspect that is becoming increasingly important as detector sensitivity improves and the catalogue of observations grows in size. Invariably, some physics must be neglected in order to achieve an adequate combination of accuracy and efficiency to enable current Bayesian inference methods to reliably recover the preferred source parameters on a reasonable timescale. 
The flagship analysis by \cite{2021arXiv211103606T} uses models by \cite{2021PhRvD.103j4056P} and \cite{2020PhRvD.102d4055O}, which include the effects of spin precession and higher-order modes but neglect eccentricity.
Conversely, waveform models that include eccentricity but neglect spins have also been used to analyse data.  Attempts in this direction include those by \cite{
2019MNRAS.490.5210R,
2020MNRAS.495..466W,
2020ApJ...903L...5R, 
2021ApJ...921L..31R,
2022arXiv220614695R,
2021arXiv210707981O,2021arXiv210605575G,2022ApJ...936..172K}
who relied on the effective-one-body approaches by \cite{2020PhRvD.101d4049L} and \cite{2020PhRvD.101j1501C}. In addition, \cite{2022arXiv220801766I} presented an analysis which include both eccentricity and higher-order modes using the approximant by \cite{2021PhRvD.103j4021N}. %
The cited eccentric models include the effects of aligned spins, but neglect spin precession.
At present, there are no readily available waveform models that can capture both spin precession and orbital eccentricity.
Completing joint parameter-estimation runs on both of these effects remains an open problem.

Despite these limitations,
several of the current events contain %
 hints of spin precession \citep{2021arXiv211103606T,2021PhRvX..11b1053A,2022Natur.610..652H,2022PhRvL.128s1102V,2022PhRvD.106j4017P} and/or eccentricity \citep{2020ApJ...903L...5R,2022arXiv220614695R,2022NatAs...6..344G,2021PhRvL.126t1101B}.  
The most emblematic event in this regard is GW190521, which is consistent with both BHs with aligned spins on eccentric orbits  \citep{2020ApJ...903L...5R} and BHs with precessing spins on quasi-circular orbits \citep{2020PhRvL.125j1102A}.  
Spot checks against numerical relativity simulations containing both eccentricity and spin precession indicate that a combination of the two could also fit the data well \citep{2022NatAs...6..344G}. 
Moreover, the same event was claimed to also be compatible with an hyperbolic encounter \citep{2021arXiv210605575G} as well as mergers of exotic compact objects \citep{2021PhRvL.126h1101B}.

Crucially, GW190521 is a short signal.
During its $\ssim 0.1$ seconds in band, the signal underwent only $\sim 5$ GW cycles, which mostly originate from the merger of the binary and the ringdown of the remnant \citep{2020PhRvL.125j1102A}.
To some extent, the ambiguity surrounding the origin of GW190521 reflects one's intuition: shorter signals are less informative and can be fitted near-equally well under a variety of different assumptions. 
In this work, we attempt to put this statement on solid footing and investigate how the 
distinguishability between eccentricity and spin precession depends on the number of GW cycles in band.

Eccentricity and spin precession share some similarities in their influence on the waveform.
In both cases, signal modulation happens on a timescale that is longer than that of the orbit, but shorter than that of the inspiral. 
In the spin precessing case, the intermediate timescale is set by the change of orientation of the orbital plane.
In the eccentric case, one must consider the timescales associated with pericenter and apocenter passages. %
In general, for the two effects to be distinguishable, the signal under analysis must have a duration that spans an appreciable portion of the added timescale.

This work is organised as follows.
In Sec.~\ref{sec:method} we describe the underlying Bayesian inference framework and the adopted post-processing strategies. 
We present our results in Sec.~\ref{sec:results}, where we show that the distinguishability of eccentricity from precession indeed increases with the length of the signal. 
It is easier to distinguish eccentric signals from precessing signals when the system has a higher eccentricity close to merger. 
For quasi-circular precessing systems, it is instead easier to distinguish precession from eccentricity when the source binary is close to edge-on and maximally precessing, even when there are very few cycles in band. 
We demonstrate that the inclusion of aligned or anti-aligned spins complicates the interpretation of the signal due to their influence on the duration of the signal: an aligned-spin system with the same measured eccentricity as an anti-aligned-spin system will in fact have a lower eccentricity at a fixed number of cycles before merger.
We conclude with a discussion and a short summary of our findings %
in Sec.~\ref{sec:conclusion}.

\section{Method}
\label{sec:method}

\subsection{Simulated sources}
\label{simsources}

We simulate the detection of signals from either eccentric or spin precessing BBH with a design-sensitivity three-detector network comprising the LIGO Hanford, LIGO Livingston,
 and Virgo instruments.
We perform parameter estimation using both the eccentric waveform model \textsc{SEOBNRE} \citep{2020PhRvD.101d4049L} and the spin-precessing waveform model \textsc{IMRPhenomPv2} \citep{2014PhRvL.113o1101H}.
\textsc{SEOBNRE} includes eccentricity and aligned spins up to an effective spin parameter $\chi_\mathrm{eff} = 0.6$, but it does not capture precession effects induced by misaligned spins.
Conversely, \textsc{IMRPhenomPv2} includes precessing spins with any orientations, but is restricted to quasi-circular sources.
\new{Neither waveform includes higher-order modes; since we only consider equal-mass sources here, the potential impact of higher modes in the signals we inject should be minimal \citep[e.g.,][]{2021PhRvD.103b4042M}.}%

It is desirable to compare sources with different number of orbital cycles in band while maintaining a constant signal amplitude. This can be achieved by varying the total mass $M$ and luminosity distance $d_\mathrm{L}$ to the source as follows
\begin{align}
    M' &= M \mathcal{F},  \\
    d_\mathrm{L}' & = d_\mathrm{L} \mathcal{F}, 
\end{align}
where $\mathcal{F}$  is a dimensionless constant.
This is conceptually similar to redshifting the source (with redshift $z=\mathcal{F}-1$), though we do not refer to this transformation as such because the adopted scale of $\mathcal{F}$ is too high.
For each set of injections described below, we use $\mathcal{F} \in [5.0, 3.0, 2.0, 1.5]$, and add $\mathcal{F} \in [1.25, 1.0]$ if the eccentricity of the injected waveform is $e_\mathrm{10~Hz} \leq 0.3$.

For eccentric sources, we convert  the reference frequency $f_\mathrm{ref}= 10$~Hz at which the eccentricity is defined,
\begin{align}
    \label{eq:freqshift}
    f_\mathrm{ref}' = \frac{f_\mathrm{ref}}{\mathcal{F}},
\end{align}
and evolve the eccentricity from $f_\mathrm{ref}'$ to $f_\mathrm{ref}$ using \citeauthor{1964PhRv..136.1224P}' (\citeyear{1964PhRv..136.1224P}) equations.
We note those are derived for binaries with non-spinning BHs and, as a result, the evolved eccentricities do not include higher post-Newtonian (PN) order terms where spins might play a role 
\citep{1995PhRvD..52..821K}.
Using the formulae presented by \cite{2018PhRvD..98j4043K} up to 1.5PN order ($n=3$) and including only up to terms up to $\mathcal{O}{(e^2)}$, we evolve the eccentricity of our ``\textit{Eccentric, aligned spins}'' series (see below), which start with $e_\mathrm{10~Hz}=0.2$ at $\mathcal{F}=1$.
For the case that requires the longest evolution and the largest change in eccentricity ($\mathcal{F}=5$), the positively (negatively) aligned spin case yields an eccentricity difference of $+0.015$ ($-0.011$) from the Peters' estimate.
When we instead evolve a non-spinning system, the difference in eccentricity is $+0.010$.
Therefore, for the systems considered here, the error introduced using the Peters' equations for spinning systems is on the order of errors on the same estimates for non-spinning systems.
Furthermore, since this is below the threshold for detectable eccentricity (e.g.,~\citealt{Lower:2018:eccentricity}), this is highly unlikely to have a significant effect on our results.

We make us use of two complementary references for the orbital eccentricity:
\begin{enumerate}[leftmargin=*] \item The quantity $e_\mathrm{10~Hz}$ indicates the eccentricity at a reference GW frequency of $10$~Hz, the parameter that is most commonly reported in current studies of BBH eccentricity.
\item The quantity $e_\mathrm{13~cycles}$ is the eccentricity measured $13$ orbital cycles before merger. 
\end{enumerate}
When shifting $M$, $d_\mathrm{L}$, and $e_\mathrm{10~Hz}$ using Eq. (\ref{eq:freqshift}) and Peters' equations, we produce a waveform that has about the same $e_\mathrm{13~cycles}$ as the unscaled set of parameters.

For all injected waveforms, we choose GW190521-like values for right ascension $\alpha = 3.3$, declination $\delta = 0.5$, phase $\phi=6.2$, polarisation  $\psi=1.6$, and geocentric time $t_{\mathrm{geo}}=1242442967.46$~s.
We vary the angle between the total angular momentum and the line of sight between three possible values: $\theta_{JN}=\pi/10$ (which is similar to that of GW190521), $\pi/4$, and $\pi/2$ (i.e., edge on).
We consider component masses $m_1=m_2=20$~M$_\odot$ when $\mathcal{F}=1$.
For each set of injections, we vary $d_\mathrm{L}$ at $\mathcal{F}=1$ so that the optimal signal-to-noise ratio of the injection is $\rho_\mathrm{opt} \simeq 25$.

We present three injection series:
\begin{enumerate}[leftmargin=*]
    \item \textit{Eccentric, non-spinning.} These signals are generated using \textsc{SEOBNRE}. We consider five sets of injections of this flavour. For three of these sets, we set $e_\mathrm{13~cycles}= 0.58,0.35,0.14$ and transform the mass, distance and eccentricity following the procedure outlined above. For the fourth set, we transform mass and distance, but keep $e_\mathrm{10~Hz}=0.10$ fixed. In both cases, we set $\theta_{JN}=\pi/10$. For the fifth set, we repeat our  $e_\mathrm{13~cycles} = 0.14$ series but change the inclination to $\theta_{JN}=\pi/2$.
    \item \textit{Eccentric, aligned spins.} These waveforms are also generated using \textsc{SEOBNRE}. We consider four sets of injections in this category, all with aligned-spin magnitudes $\chi_1=\chi_2=0.59$, two with component spin tilt angles $\theta_1=\theta_2=0$ (aligned spin), and two with $\theta_1=\theta_2=\pi$ (anti-aligned spin). One of each aligned and anti-aligned subset is edge-on, while the other has $\theta_{JN} = \pi/10$. These runs all have $e_\mathrm{10~Hz} = 0.20$ when $\mathcal{F}=1$.
    \item \textit{Quasi-circular, precessing spins.} These waveforms are generated using \textsc{IMRPhenomPv2}. We inject simulated waveforms from three highly spin precessing systems with $\chi_1=\chi_2=0.99$, $\theta_1=\theta_2=\pi/2$, angle between the azimuth angles of the spin vectors on the orbital plane $\phi_{12}=\pi$, and angular difference between the orbital and total angular momenta azimuths $\phi_{JL}=\pi$. The injected signals differ only in their $\theta_{JN}$ values, which are $\pi/10, \pi/4$ and $\pi/2$. All spin-dependent quantities are quoted at a reference GW frequency of $10$~Hz.
\end{enumerate}

For each injected waveform we count the number of orbital cycles in band.
This is done by first extracting the 
the (unwrapped) phase $\phi_\mathrm{gw}$ of the waveform
\begin{equation}
    \phi_\mathrm{gw} = \arctan{\frac{h_\times}{h_+}}, 
\end{equation}
where $h_+$ and $h_\times$ are the two GW polarisations, 
and hence the evolution of the GW-frequency
\begin{equation}
    f_\mathrm{gw} = \frac{1}{\pi}\frac{d\phi_\mathrm{gw}}{d t}.
\end{equation}
For quasi-circular sources, the number of orbital cycles is given by \citep{2014LRR....17....2B}
\begin{equation}
    N_\mathrm{cycles} = \frac{\phi_\mathrm{gw}^\mathrm{ISCO} - \phi_\mathrm{gw}^{f_0}}{2 \pi},
\end{equation}
where $\phi_\mathrm{gw}^\mathrm{ISCO}$ is the phase of the waveform as it reaches the innermost stable circular orbit (ISCO) and $\phi_\mathrm{gw}^{f_0}$ is the phase of the waveform when it reaches the detector.
For eccentric waveforms, we visually inspect the time-dependent frequency evolution $f_\mathrm{gw}(t)$ and count the number of apastron passages (represented as peaks in the frequency evolution) before the plunge above our chosen minimum analysis frequency of $f_0=20$~Hz.

We perform Bayesian parameter estimation using \textsc{Bilby} \citep{2019ApJS..241...27A}. 
We consider data segments of $8$~s and a sampling frequency of $4096$~Hz.
We generate each waveform from $10$~Hz, set a minimum analysis frequency of $20$~Hz and a maximum frequency of $2048$~Hz.
When running the analysis, we marginalize over both phase and time. We use the \textsc{dynesty} sampler \citep{2020MNRAS.493.3132S} with its \textsc{Bilby} default  settings.

We use uninformative priors as commonly used in GW astronomy (cf. \citealt{2019PhRvX...9c1040A,2021arXiv211103606T,2021PhRvX..11b1053A}). %
When sampling in component aligned spins $\chi_i \mathrm{cos}(\theta_i)$, we restrict the magnitude to $\chi_i \leq0.59$, to remain within the validity limits of \textsc{SEOBNRE}.
When sampling over eccentricity, we use a prior that is log-uniform over $e_\mathrm{10~Hz}\in [10^{-4}, 0.2]$, unless the injected waveform has $e_\mathrm{10~Hz} > 0.2$, in which case we raise the upper limit to a maximum of $e_\mathrm{10~Hz} = 0.30$.
Changing prior limits impacts the Bayes factor (specifically, the Occam penalty is more severe for an analysis with a larger prior volume).
To account for this, we (i) offset Bayes factors calculated using a wider eccentricity prior by the difference in volume between a log-uniform prior with $e_{10~\mathrm{Hz},\mathrm{max}} = 0.30$ and $0.20$ and (ii) offset Bayes factors calculated using a wider spin-magnitude prior by the difference in volume between a uniform prior with $\chi_{i, \mathrm{max}}=0.99$ and $0.6$.
The difference is small in both cases (ln~$\Delta\pi_{e_{10~\mathrm{Hz}}} = -0.35$, ln~$\Delta\pi_{\chi_i} = -0.50$).

\subsection{Model-selection strategies}

Eccentricity inference is performed using the likelihood-reweighting procedure introduced by \cite{2019PhRvD.100l3017P}. %
The high computational cost of \textsc{SEOBNRE} is prohibitive for direct Bayesian inference schemes. Despite recent advances in parallelization of GW pipelines \citep{2018arXiv180510457L,2020MNRAS.498.4492S}, %
full \textsc{SEOBNRE}  runs
still require drastically reduced priors and extreme computing resources \citep{2021ApJ...921L..31R} which is not practical for the relatively large parameter-space exploration demonstrated here. 

The broad strategy is to estimate the posterior probability distribution that would be obtained with a computationally-inefficient ``target'' model by first exploring the parameter space with a more efficient ``proposal'' model. For our proposals we use the quasi-circular, aligned-spin waveform model \textsc{IMRPhenomD} \citep{2016PhRvD..93d4007K} because it is fast to generate, has been proven to facilitate the recovery of low-to-moderate eccentricities via likelihood reweighting \citep{2019MNRAS.490.5210R,2020ApJ...903L...5R}, and has been shown to enable results %
that are consistent with those obtained via direct parameter estimation \citep{2021ApJ...921L..31R}.

Our goal is to approximante the target (eccentric) posterior probability distribution $p_\mathrm{E}(\theta|d)$, where $\theta$ is a vector of parameters and $d$ are the analysed data.
We first obtain the proposal posterior distribution $p_\mathrm{C}(\theta|d)$ and transform this into $p_\mathrm{E}(\theta|d)$ by evaluating the ratio of the target and proposal likelihoods $\mathcal{L}_\mathrm{E}(d|\theta)$ and $\mathcal{L}_\mathrm{C}(d|\theta)$ for each posterior sample
\begin{align}
    p_\mathrm{E}(\theta|d) &= p_\mathrm{C}(\theta|d) \times \frac{\mathcal{L}_\mathrm{E}(d|\theta)}{\mathcal{L}_\mathrm{C}(d|\theta)} \times \frac{\mathcal{Z}_\mathrm{C}}{\mathcal{Z}_\mathrm{E}}\,.
\end{align}
Here, ${\mathcal{Z}_\mathrm{C}}/{\mathcal{Z}_\mathrm{E}}$ is the ratio of model evidences\new{, and $\mathcal{L}_\mathrm{E}$ is the likelihood using the eccentric model and marginalising over eccentricity}. The inverse of this is the Bayes factor in favour of the eccentric hypothesis over the quasi-circular hypothesis, which can be calculated using \citep{2019PhRvD.100l3017P}

\begin{align}
    \mathcal{B}_{\mathrm{E}/\mathrm{C}} = \frac{\mathcal{Z}_\mathrm{E}}{\mathcal{Z}_\mathrm{C}} = \frac{1}{N} \sum_i^N{    \frac{\mathcal{L}_\mathrm{E}(d|\theta_i)}{\mathcal{L}_\mathrm{C}(d|\theta_i)} },
\end{align}
where $i$ labels the posterior samples and $N$ is their total number. The Bayes factor between the eccentric and spin precessing hypothesis, $\mathcal{B}_{\mathrm{E}/\mathrm{P}}$, can then be calculated by multiplying $\mathcal{B}_{\mathrm{E}/\mathrm{C}}$ by $\mathcal{B}_{\mathrm{C}/\mathrm{P}}$, where $\mathcal{B}_{\mathrm{C}}$ and $\mathcal{B}_{\mathrm{P}}$ are outputs of the \textsc{Bilby} analyses with \textsc{IMRPhenomD} and \textsc{IMRPhenomPv2}, respectively.

As with all reweighting strategies, this method can only be utilized when the proposal and target distribution are sufficiently similar, i.e. $p_\mathrm{C}(\theta|d) \approx p_\mathrm{E}(\theta|d)$. 
For some of our runs, namely those with large aligned spins and eccentricity, we find that this is not the case.
Eccentricity and aligned spins have opposite influences on the duration of the waveform and cause the orbital frequency to oscillate differently than in the non-spinning or negatively aligned-spin case. 
In the maximally aligned-spin case, the local maxima and minima of the frequency evolution with time do not map to periastron and apastron passages.
Because the inclusion of eccentricity complicates the aligned-spin waveform significantly, the quasi-circular analysis finds that the most preferred region of parameter space is significantly removed from the injected value, signalling an evident problem for the reweighing procedure. 

As an alternative measure of distinguishability for those injections, we compute the quantity 
\begin{equation}
    \rho = \rho_\mathrm{mf}\rho_\mathrm{opt} - \frac{1}{2}\rho^2_\mathrm{opt},
   \label{rhoproxy}
\end{equation}
where $\rho_\mathrm{mf}$ is the matched-filter signal-to-noise ratio (SNR) and $\rho_\mathrm{opt}$ is the optimal SNR.
We determine the fractional difference between $\rho_\mathrm{inj}$, the quantity above evaluated for the injected signal, and $\Bar\rho_\mathrm{p}$, the integral of $\rho$ calculated over the posterior recovered with the spin precessing model.
This relates to the likelihood, since ln~$\mathcal{L}$ can be written as %
\citep{2019PASA...36...10T}
\begin{equation}
    \mathrm{ln}~\mathcal{L} = C + \rho_\mathrm{mf}\rho_\mathrm{opt} - \frac{1}{2}\rho^2_\mathrm{opt}.
\end{equation}
Here $C$ is a factor proportional to the noise log likelihood, which is the same for any analysis performed on the same data with the same sampler settings.

\begin{figure*}
    \centering
\includegraphics[width=0.7\textwidth]{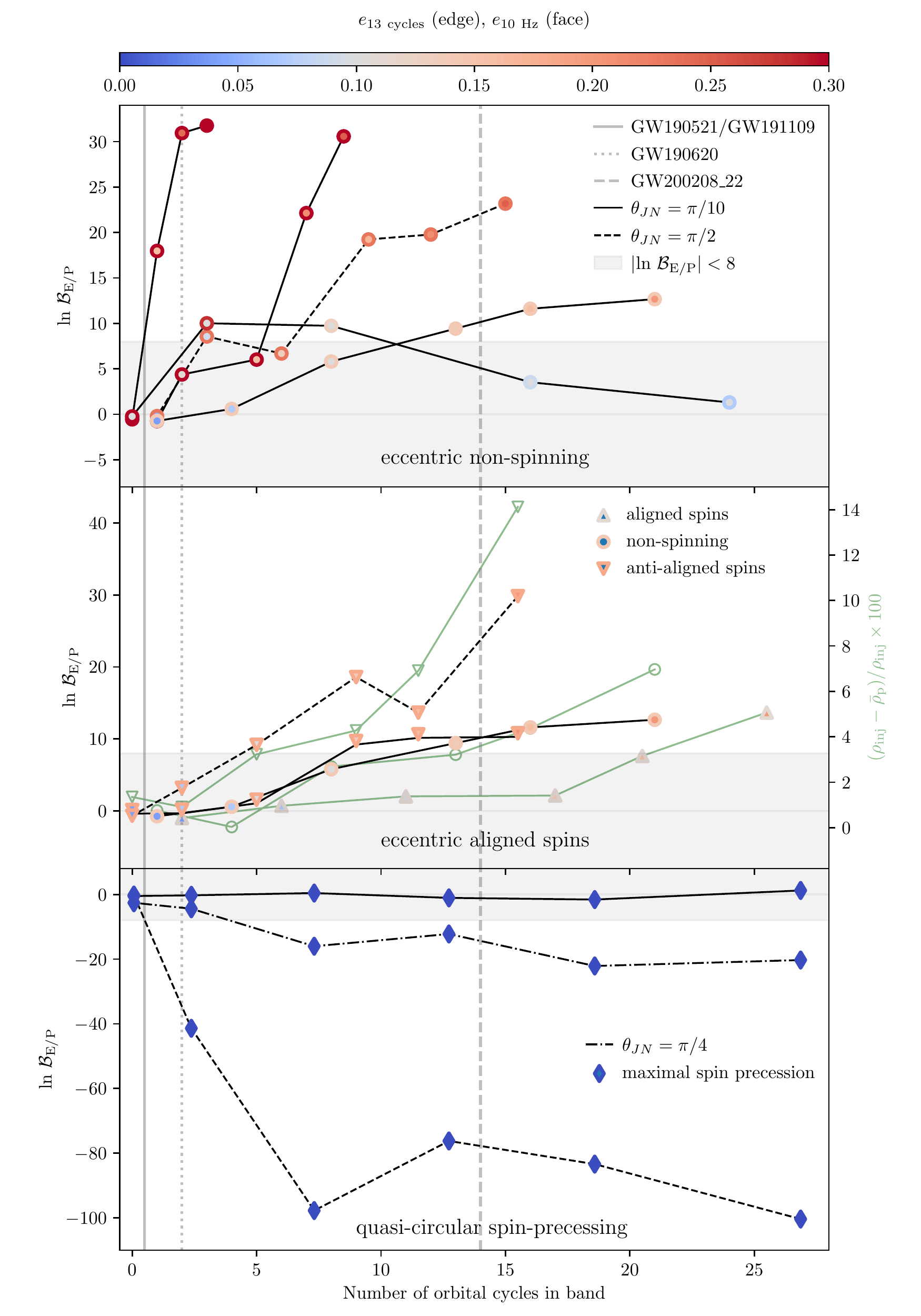}
    \caption{Natural log Bayes factor of eccentricity vs. spin precession as a function of the number of orbital cycles visible in the inspiral. Positive (negative) values of  $\ln \mathcal{B}_{\mathrm{E/P}}$ indicate a preference for the aligned-spin eccentric (spin precessing quasi-circular) model. The three panels contain results for each of the injection series described in Sec.~\ref{simsources}. The top panel shows results for eccentric and non-spinning injections, the middle panel shows results for eccentric and spin-aligned injections, and the bottom panel shows results for  quasi-circular and spin precessing injections.
    For each run, the value of the detector-frame eccentricity $e_\mathrm{10~Hz}$ is indicated by the face colour of the marker and the corresponding source-frame eccentricity $e_{\mathrm{13~cycles}}$ is indicated by the edge colour. Each marker is linked to others in the same injection subset with a grey line. The linestyle indicates the inclination of the source; solid, dot-dashed, and dashed lines are used for $\theta_{JN}=\pi/10$ (i.e. similar to that of GW190521), $\theta_{JN}=\pi/4$, and $\theta_{JN}=\pi/2$ (i.e. edge-on), respectively.
    The significance region with $|\ln \mathcal{B}_\mathrm{E/P}| < 8$, a conventional value for establishing confidence that one model is preferred over another, is indicated with grey shading.
    The approximate number of orbital cycles in band for four eccentric-event candidates are indicated with vertical light-grey lines. 
  For injections with eccentricity and aligned spins (middle panel) we pair the Bayes factors $\ln \mathcal{B}_{\mathrm{E/P}}$ (left vertical scale in black) to the approximate criterion based on $\rho$ from Eq.~(\ref{rhoproxy}) (right vertical scale in green).
  For injection series with equivalent $\ln\mathcal{B}_\mathrm{E/P}$ values in this panel, we plot $\rho$ with unfilled markers to avoid overcomplicating the plot.
  We do not plot the fractional change in $\rho$ for edge-on systems since the results are very similar to those already shown for close to face-on injections. %
      \label{fig:cycles_vs_bayes_all}}
\end{figure*}

\section{Results}
\label{sec:results}
Our results are presented in Fig.~\ref{fig:cycles_vs_bayes_all}, where we show the number of orbital cycles in band versus the resulting Bayes factor $\ln \mathcal{B}_{\mathrm{E/P}}$ for the aligned-spin eccentric model against the spin precessing quasi-circular model.
For context, we also show the number of cycles of four events that have previously been flagged as potential candidates for containing the signatures of eccentricity: GW190521, GW190620, GW191109, and GW200208\_22 \citep{2022arXiv220614695R}.
The number of in-band orbital cycles for these systems were \new{estimated} by taking the maximum \emph{a posteriori} from the cited analysis and counting the number of apastron passages visible before the plunge in the time-frequency evolution plot.

The top panel of Fig~ \ref{fig:cycles_vs_bayes_all} shows results for our ``\textit{Eccentric, non-spinning}'' injection series. 
The broad trend is that the distinguishability between precession and eccentricity increases with the length of the signal.  
It is also not possible to distinguish a preference when either eccentricity or spin precession have a small effect on the signal, for example when the eccentricity is small or a spin precessing binary has a mostly face-on ($\theta_{JN} = \pi/10$) orientation. %
When fewer orbital cycles are visible, the same value of $e_{\mathrm{13~cycles}}$ corresponds to a lower value of  $e_\mathrm{10~Hz}$, so the eccentric model cannot be distinguished as a better fit to the data. 
When only the merger is visible in band (i.e. number of orbital cycles is $\ssim 0$), it is not possible to distinguish a preference between the models ($\ln \mathcal{B}_{\mathrm{E/P}}\ssim 0$). 
We find that $\mathcal{B}_{\mathrm{E/P}}$ is strictly increasing for the four injection series where we fix $e_{\mathrm{13~cycles}}$. 
The non-monotonic behaviour of the Bayes factor for the series where $e_\mathrm{10~Hz}$ is kept fixed is due to the fact that, when many orbital cycles are in band, the detector-frame eccentricity $e_\mathrm{10~Hz}$ corresponds to a lower source-frame eccentricity $e_{\mathrm{13~cycles}}$.
Overall, we find that higher source-frame eccentricity $e_{\mathrm{13~cycles}}$ and more cycles in band both make eccentricity easier to distinguish from spin precession. 
Indeed, the two effects go hand-in-hand: for our set of parameters, a signal with 8 cycles in band has almost exactly the same $\ln\mathcal{B}_{\mathrm{E/P}}$ as a signal with 4 cycles in band when $e_\mathrm{10~Hz}$ is fixed.

Compared to their non-spinning counterparts, BBHs with aligned (anti-aligned) spins present a larger (smaller) number of orbital cycles~\citep{2006PhRvD..74d1501C}.
This, in turn, impacts the eccentricity/precession distinguishability. 
This is demonstrated in the middle panel of Fig.~ \ref{fig:cycles_vs_bayes_all}, where we show results for our ``\textit{Eccentric, aligned spins''} injection series.
A given value of $e_\mathrm{10~Hz}$ corresponds to a lower (higher) value of $e_{\mathrm{13~cycles}}$ for a binary with aligned (anti-aligned) spins.
As a result, longer signals are required to distinguish eccentricity and spin precession for an aligned-spin system compared to sources with anti-aligned spins measured with the same $e_\mathrm{10~Hz}$. 
We plot in this central panel both $\ln\mathcal{B}_{\mathrm{E/P}}$ and the percentage difference between $\Bar{\rho}_\mathrm{p}$ and $\rho_\mathrm{inj}$.
For runs with zero and aligned spins, we plot both distinguishability measures, to demonstrate that the two measures exhibit similar trends.
The two measures both support the conclusion that it is easier to confidently distinguish eccentricity from spin precession in longer waveforms.
Since $\rho$ is less sensitive to changes in $\theta_{JN}$, we leave out the results of runs with $\theta_{JN}=\pi/2$ in the panel for ease of readability, but note that they trace their $\theta_{JN}=\pi/10$ counterparts closely.

Finally, the bottom panel of Fig.~\ref{fig:cycles_vs_bayes_all} illustrates the case where an injected precessing signal is analysed with an eccentric model (``\textit{Quasi-circular, precessing spins}'' series).
In this case, we find that  the inclination of the source is a dominant factor to determine the distinguishability of spin precession and eccentricity.
This is because the influence of spin precession on the signal will be most pronounced when the binary is edge-on to the observer \citep{1994PhRvD..49.6274A}.
When a system is maximally spin precessing but close to face-on, as is the case for GW190521 \citep{2020PhRvL.125j1102A}, the Bayes factor between the eccentric and spin precessing hypotheses will not be compelling.
When a system is instead closer to edge-on (for the same SNR, of course), the signs of spin precession in the signal are stronger, and hence easier to distinguish from those of eccentricity. For our injections with $\theta_{JN} = \pi/2$, we observe a transition in $\mathcal{B}_{\mathrm{E/P}}$ when the number of visible orbital cycles is $\gtrsim 5$. While the specific threshold will depend on the injected parameters, we believe this trend to be sufficiently generic, as the distinguishability increases as soon as the signal covers a (portion of a) spin precession period.  

\section{Discussion \& Conclusion}
\label{sec:conclusion}

We show explicitly that, in accordance with our intuition, GW signals need to \new{be} long enough before one can tell spin precession and eccentricity apart. We quantify this in terms of the number of visible orbits. Some favourable configurations, notably including high eccentricities and edge-on highly precessing systems, can instead be confidently identified with just a few orbital cycles in band.
This implies that, while eccentricity and spin precession can indeed be confused for extremely short signals like GW190521, the two effects do not induce a genuine observational degeneracy.
\new{Additionally, by comparing the number of orbital cycles visible for detected eccentric candidates, we infer both signals GW190620 and GW200208\_22 are sufficiently long that strong spin precession or eccentricity signatures should be distinguishable.
This implies that any such signature is low in these signals since ln~$\mathcal{B}_{\mathrm{E/P}}<8$ in reality \citep{2022arXiv220614695R}.
Meanwhile, smaller $\mathcal{B}_{\mathrm{E/P}}$ values are obtained for highly eccentric or spin-precessing injections with similar lengths to GW190521 and GW191109, indicating that even a strong signature of eccentricity or spin-induced precession would be indistinguishable for these sources. }

Our key findings are as follows:

\begin{itemize}
\item When the detector-frame eccentricity $e_{\mathrm{13~cycles}}$ is kept fixed, the Bayes factor $\mathcal{B}_{\mathrm{E/P}}$ increases as the number of cycles in band increases.
\item %
Systems with a positive (negative) aligned spins will spend a longer (shorter) time in band. Consequently, we find that eccentricity is harder to distinguish from precession for system with positive aligned spins and easier for those with negative aligned spins. %
\item It is not possible to confirm a strong preference for spin precession in GW data when a precessing BBH is observed face-on with respect to the observer. 
\item On the other hand, the spin precessing hypothesis can be strongly preferred over the eccentric hypothesis for spin precessing systems with a larger inclination, even with $< 5$ orbital cycles in band.
\end{itemize}

While correlations between $e_\mathrm{10~Hz}$ and the full parameter space of spin magnitudes and tilt angles are yet to be investigated, our results suggest that, when found, strong evidence (i.e., a large $|\mathrm{ln}~\mathcal{B}_\mathrm{E/P}|$)
in favour of either effect should be considered robust.

\section*{Acknowledgements}

We thank Johan Samsing and the organizers of the 2022 Neils Bohr International Academy Workshop on Black-Hole Dynamics in Copenhagen where this work was initiated. %
We thank Juan Calder\'on Bustillo, Uli Sperhake, Shichao Wu, Daria Gangardt,  Matthew Mould, and Rossella Gamba for discussions.
I.M.R.-S. acknowledges support received from the Herchel Smith Postdoctoral Fellowship Fund and the OzGrav Travel Award Scheme.
D.G. is supported by Leverhulme Trust Grant No. RPG-2019-350, European Union's H2020 ERC Starting Grant No. 945155--GWmining, and Cariplo Foundation Grant No. 2021-0555.
N.L. is supported by European Union's H2020 ERC Starting Grant agreement No. 757480--DarkGRA, MIUR PRIN and FARE programs No.~B84I20000100001---GWNext, and the Amaldi Research Center funded by MIUR program ``Dipartimento di Eccellenza'' No.~B81I18001170001. Computational work was performed at CINECA with allocations through INFN, Bicocca, and ISCRA project HP10BEQ9JB.
The authors are grateful for computational resources provided by the LIGO Laboratory and supported by National Science Foundation Grants PHY-0757058 and PHY-0823459.

\section*{Data Availability}
The data underlying this article will be shared on reasonable request to the corresponding author.

\bibliographystyle{mnras_tex_edited}
\bibliography{prec_or_ecc}

\end{document}